\documentclass{ieeetj}
\usepackage{cite}
\usepackage{amsmath,amssymb,amsfonts}
\usepackage{algorithmic}
\usepackage{graphicx,color}
\usepackage{textcomp}
\usepackage{xcolor}
\usepackage{hyperref}
\usepackage{multirow}
\usepackage{makecell}
\hypersetup{hidelinks=true}
\usepackage{algorithm,algorithmic}
\def\BibTeX{{\rm B\kern-.05em{\sc i\kern-.025em b}\kern-.08em
    T\kern-.1667em\lower.7ex\hbox{E}\kern-.125emX}}
\AtBeginDocument{\definecolor{tmlcncolor}{cmyk}{0.93,0.59,0.15,0.02}\definecolor{NavyBlue}{RGB}{0,86,125}}

\newcommand\blfootnote[1]{%
  \begingroup
  \renewcommand\thefootnote{}\footnote{#1}%
  \addtocounter{footnote}{-1}%
  \endgroup
}

\def\authorrefmark#1{\ensuremath{^{\textbf{#1}}}}

\begin{document}

\markboth{Objective Evaluation of Prosody and
Intelligibility in Speech Synthesis via
Conditional Prediction of Discrete
Tokens}{Ulgen {et al.}}

\title{Objective Evaluation of Prosody and Intelligibility in Speech Synthesis via Conditional Prediction of Discrete Tokens}

\author{ISMAIL RASIM ULGEN\authorrefmark{1}, ZONGYANG DU\authorrefmark{1}, JUNCHEN LU\authorrefmark{2}, PHILIPP KOEHN\authorrefmark{1}, \\ BERRAK SISMAN\authorrefmark{1}}
\affil{Center for Language and Speech Processing (CLSP), Johns Hopkins University, Baltimore, MD, USA}
\affil{National University of Singapore, Singapore}
\authornote{This work is supported by NSF CAREER award IIS-2533652.}

\begin{abstract}
Objective evaluation of synthesized speech is critical for advancing speech generation systems, yet existing metrics for intelligibility and prosody remain limited in scope and weakly correlated with human perception. Word Error Rate (WER) provides only a coarse text-based measure of intelligibility, while F0-RMSE and related pitch-based metrics offer a narrow, reference-dependent view of prosody. To address these limitations, we propose \textit{TTScore}, a targeted and reference-free evaluation framework based on conditional prediction of discrete speech tokens. TTScore employs two sequence-to-sequence predictors conditioned on input text: \textit{TTScore-int}, which measures intelligibility through content tokens, and \textit{TTScore-pro}, which evaluates prosody from the perspective of pitch, through prosody tokens. For each synthesized utterance, the predictors compute the likelihood of the corresponding token sequences, yielding interpretable scores that capture alignment with intended linguistic content and prosodic structure. Experiments on the SOMOS, VoiceMOS, and TTSArena benchmarks demonstrate that TTScore-int and TTScore-pro provide reliable, aspect-specific evaluation and achieve stronger correlations with human judgments of overall quality than existing intelligibility and prosody-focused metrics.

\end{abstract}



\maketitle

\section{INTRODUCTION}

Speech is a natural and effective form of human communication. Building systems that generate high-quality, natural-sounding speech is a central goal of speech synthesis \cite{ttssurvey,sisman_overview}. Evaluating these systems is equally important, as it shows whether advances in modeling improve perception. 
\blfootnote{\footnotesize{Code and pre-trained models: https://github.com/rsmlgen/TTScore}}

Traditionally, evaluation has relied on subjective and objective methods \cite{yamagishi_overview}. In subjective tests, listeners rate the naturalness or overall quality of speech. While these tests directly capture human perception, they are costly, time-consuming, and difficult to scale. Objective evaluation offers automatic and inexpensive alternatives, but typically focuses on narrower aspects such as intelligibility\cite{diagnostics} or prosody \cite{oytun_overview}. This creates a gap: subjective evaluation reflects overall quality, whereas objective metrics capture isolated properties. For evaluation to be reliable, aspect-specific metrics must not only measure the targeted dimension but also align with broader human judgments \cite{das20_vccbc}. In this work, we address this need by proposing objective metrics for intelligibility and prosody that remain targeted yet show stronger alignment with overall naturalness as perceived by human listeners.

Two core aspects strongly influence human judgments of naturalness and quality: intelligibility and prosody. Intelligibility reflects how well the linguistic content is conveyed in speech, directly affecting usability across domains such as assistive technologies, human-computer dialogue systems, and machine translation \cite{sking_refining}. Even highly natural-sounding speech becomes ineffective if it is not intelligible. Prosody, on the other hand, encodes rhythm, stress, and intonation, giving speech its expressiveness and supporting the communication of pragmatic and emotional meaning~\cite{prosody_reviewe}. Appropriate prosody is critical for engaging and effective human--computer interaction. Despite their importance, the objective evaluation of intelligibility and prosody in synthesized speech remains an open challenge.  

For intelligibility, widely used objective metrics such as Word Error Rate (WER) and Character Error Rate (CER) measure the discrepancy between automatic speech recognition (ASR) transcriptions and ground-truth text~\cite{yamagishi_overview,diagnostics}. While simple and effective at a coarse level, these metrics operate in the text domain and fail to capture acoustic and temporal artifacts that influence human perception. Furthermore, as both speech synthesis and ASR systems improve, reported WER and CER values continue to decrease, making them less discriminative and informative for comparing modern systems~\cite{kim2023pflow,valle,xtts}.

For prosody, objective evaluation has been even more limited. Metrics such as F0 root-mean-square error (F0-RMSE) ~\cite{Clark1999ObjectiveMF} and pitch correlation \cite{hermes} remain common \cite{Li2022FreevcTH,diffprosody,lee2022hierspeech}. However, these metrics face several drawbacks: they typically require a reference utterance, are sensitive to alignment errors, and ignore the sequential structure of prosody~\cite{sisman_overview}. Importantly, they often correlate poorly with subjective judgments of naturalness~\cite{yamagishi_overview}. As a result, subjective listening tests remain the dominant method for prosody evaluation, despite their cost and inefficiency.

These limitations highlight the need for more reliable, interpretable, and aspect-focused objective metrics. Recent advances in self-supervised learning and speech coding have made it possible to represent speech as discrete tokens that capture different aspects of the signal. For example, discrete units derived from HuBERT have been shown to correlate strongly with phoneme classes and encode fine-grained linguistic information~\cite{Hsu2021HuBERTSS}, while FACodec tokens can be trained to disentangle prosody from content and acoustic details~\cite{naturalspeech3}. Such representations open the door to evaluation metrics that directly target intelligibility or prosody while remaining scalable and reference-free.

A recent study, SpeechLMScore~\cite{Maiti2023SpeechlmscoreES}, showed that likelihood modeling over discrete speech tokens can provide useful signals for evaluating synthesized speech. While encouraging, SpeechLMScore was designed as a general-purpose quality metric: it uses a decoder-only language model without any conditioning, and it does not explicitly target intelligibility or prosody. As a result, it offers limited interpretability for analyzing in which aspects a system performs well or poorly. Instead, we propose that conditional likelihood, specifically the likelihood of aspect-specific speech tokens given the input text, is a more appropriate formulation. This measure allows more direct and correct formulations targeted to measure specific attributes of synthesized speech, resulting in more detailed and interpretable measures.  

In this work, we propose \textit{TTScore}, a targeted evaluation framework that formulates the assessment of intelligibility and prosody as conditional generation tasks. We introduce two sequence-to-sequence predictors:  
\begin{itemize}
    \item \textit{TTScore-int} for intelligibility, based on content tokens that capture linguistic content.  
    \item \textit{TTScore-pro} for prosody, based on prosody tokens that encode prosodic aspects of speech.  
\end{itemize}

Both predictors take input text and estimate the conditional likelihood of the corresponding discrete token sequence extracted from synthesized speech. The resulting likelihood scores reflect how well the generated speech aligns with the intended linguistic content or has an appropriate prosodic structure. Unlike existing approaches, TTScore-int and TTScore-pro operate directly in the speech domain, are reference-free in terms of ground-truth speech, and provide meaningful measures that correlate with both aspect-specific performance and overall perceived quality.  

\noindent Our contributions are summarized as follows:  
\begin{enumerate}
    \item We introduce a new paradigm for targeted evaluation of synthesized speech, where intelligibility and prosody are modeled as conditional generation tasks over discrete tokens.  
    \item We propose two novel metrics, TTScore-int and TTScore-pro, that separately evaluate intelligibility and prosody, overcoming the limitations of current objective metrics.  
    \item Through experiments on SOMOS, VoiceMOS, and TTSArena benchmarks, we demonstrate that TTScore-int and TTScore-pro not only provide meaningful, fine-grained assessments of intelligibility and prosody but also align more closely with human judgments of overall quality.  
\end{enumerate}
This paper is organized as follows: Section 2 reviews related work, highlighting the limitations of existing metrics and exploring sequence-to-sequence evaluation paradigms. Section 3 introduces the proposed metric, TTScore, detailing the text-to-speech token generator, different discrete speech tokens, and the calculation of the intelligibility and pitch-related prosody score. Section 4 describes the experimental setup, datasets used, and implementation details. Section 5 presents experiments and discussions for intelligibility evaluation. Section 6 presents experiments for prosody evaluation. Section 7 discusses the limitations and finally, Section 8 concludes the paper and discusses future directions.

\vspace{-2mm}
\section{RELATED WORK}
 
\subsection{SPEECH INTELLIGIBILITY EVALUATION}
With the advancements in robust automatic speech recognition (ASR) methods such as wav2vec 2.0~\cite{Hsu2021HuBERTSS} and Whisper~\cite{whisper}, evaluating intelligibility via ASR on synthesized speech has become standard practice~\cite{kim2023pflow,valle,xtts,Li2022FreevcTH}. This approach measures intelligibility in the textual domain by comparing recognized text with ground-truth transcriptions, typically reporting Word Error Rate (WER) or Character Error Rate (CER)~\cite{yamagishi_overview,diagnostics}. While widely used, these measures provide only coarse-grained signals and fail to capture local artifacts or acoustic details that influence perception. Moreover, as synthesis and ASR systems improve, reported WER and CER continue to decrease, reducing their discriminative ability when comparing modern systems. This motivates the need for fine-grained, speech-domain measures of intelligibility.  
\vspace{-3mm}
\subsection{SPEECH PROSODY EVALUATION}
Prosody plays a central role in speech naturalness, shaping rhythm, stress, and intonation. Despite its importance, the objective evaluation of prosody in synthesized speech has received less attention than intelligibility. Existing metrics are dominated by F0-based measures as a very fundamental aspect of prosody, such as F0 root-mean-square error (F0-RMSE) and F0 correlation~\cite{Clark1999ObjectiveMF}. However, these approaches rely on the availability of reference utterances, require frame-level alignment with reference speech, and often correlate poorly with human judgments of naturalness~\cite{sisman_overview,oytun_overview,yamagishi_overview}. Importantly, reference utterances are not always available (e.g., in cross-lingual style transfer), making such metrics impractical. In addition, synthesized speech may have multiple appropriate prosodic renditions that differ from the reference. As a result, subjective MOS ratings remain the dominant method for prosody evaluation, despite their inefficiency and high cost.
\vspace{-2mm}
\subsection{EVALUATION BY CONDITIONAL LIKELIHOOD}
Reference-guided evaluation has been studied in natural language processing (NLP), particularly in tasks such as summarization and translation \cite{becker2024textgenerationsystematicliterature}, where the consistency between a generated output and its source text is assessed. For example, BARTScore~\cite{bartscore} computes the conditional likelihood of a reference or candidate sentence given its source text, utilizing a text-generator and its task-specific knowledge. Similar likelihood-based formulations have also been investigated in speech, such as SpeechLMScore~\cite{Maiti2023SpeechlmscoreES}, which evaluates synthesized speech by the token likelihoods with a decoder-only language model. While promising to use the knowledge from generative models for evaluation, these methods are general-purpose or do not explicitly target intelligibility or prosody.

\subsection{SUMMARY OF RESEARCH GAP}
We identify three major gaps in current evaluation of speech synthesis:
\begin{itemize}
    \item Intelligibility metrics (WER, CER) are coarse and text-based. They fail to capture acoustic or temporal artifacts that affect perception and are becoming less discriminative as systems improve.
    \item Prosody metrics ($F0$-based) depend on references, are alignment-sensitive, correlate poorly with human ratings, and cannot capture multiple prosodic renditions. 
    \item Likelihood-based methods (e.g. SpeechLMScore) show promise but remain general-purpose and do not explicitly target intelligibility or prosody.
\end{itemize}
This work addresses these gaps by introducing targeted, reference-free evaluation metrics for intelligibility and prosody that better align with human perception of naturalness.
\vspace{-3mm}
\section{THE PROPOSED SCORE: TTScore}

A central question in evaluating synthesized speech is whether to rely on broad measures of overall quality or on targeted measures of specific aspects such as intelligibility and prosody. The ultimate goal of speech generation is to produce high-quality, natural-sounding output, and evaluations often rely on mean opinion scores (MOS) as a comprehensive indicator. While MOS captures overall perception, it does not reveal the specific strengths and weaknesses of a system. Intelligibility and prosody, however, are essential in their own right, as deficiencies in either directly degrade naturalness and usability. We argue that broad and targeted evaluations should be interconnected. Broad measures such as MOS provide an overall view of performance but are less interpretable for diagnosis, whereas targeted evaluations highlight specific shortcomings but may not fully represent overall quality. Ideally, targeted metrics should measure the intended aspect while also aligning with broader perceptual judgments.

 Motivated by this, we propose TTScore, fine-grained metrics for intelligibility, namely TTScore-int, and prosody, namely TTScore-pro, that operate directly in the speech domain. Both variants of TTScore are formulated as conditional generation tasks, where sequence-to-sequence (seq2seq) autoregressive models predict discrete speech token sequences from text. Each score is defined as the average log-likelihood of the synthesized tokens given the input text, computed with teacher forcing. TTScore-int evaluates intelligibility by measuring the alignment between content tokens and the input text, providing a finer-grained assessment than text-based metrics such as WER or CER. TTScore-pro evaluates prosody by focusing on pitch patterns and measuring the alignment between prosody tokens and the intended content. Unlike F0-based metrics, it does not require a reference utterance; thus, it is robust to alignment errors and captures the sequential structure of pitch. Crucially, TTScore-pro can accommodate multiple valid prosodic renditions without relying on a single reference.

\begin{figure*}
\centering
\label{fig:proposed_diagram}
\scalebox{0.87}{
    \includegraphics[width=\linewidth]{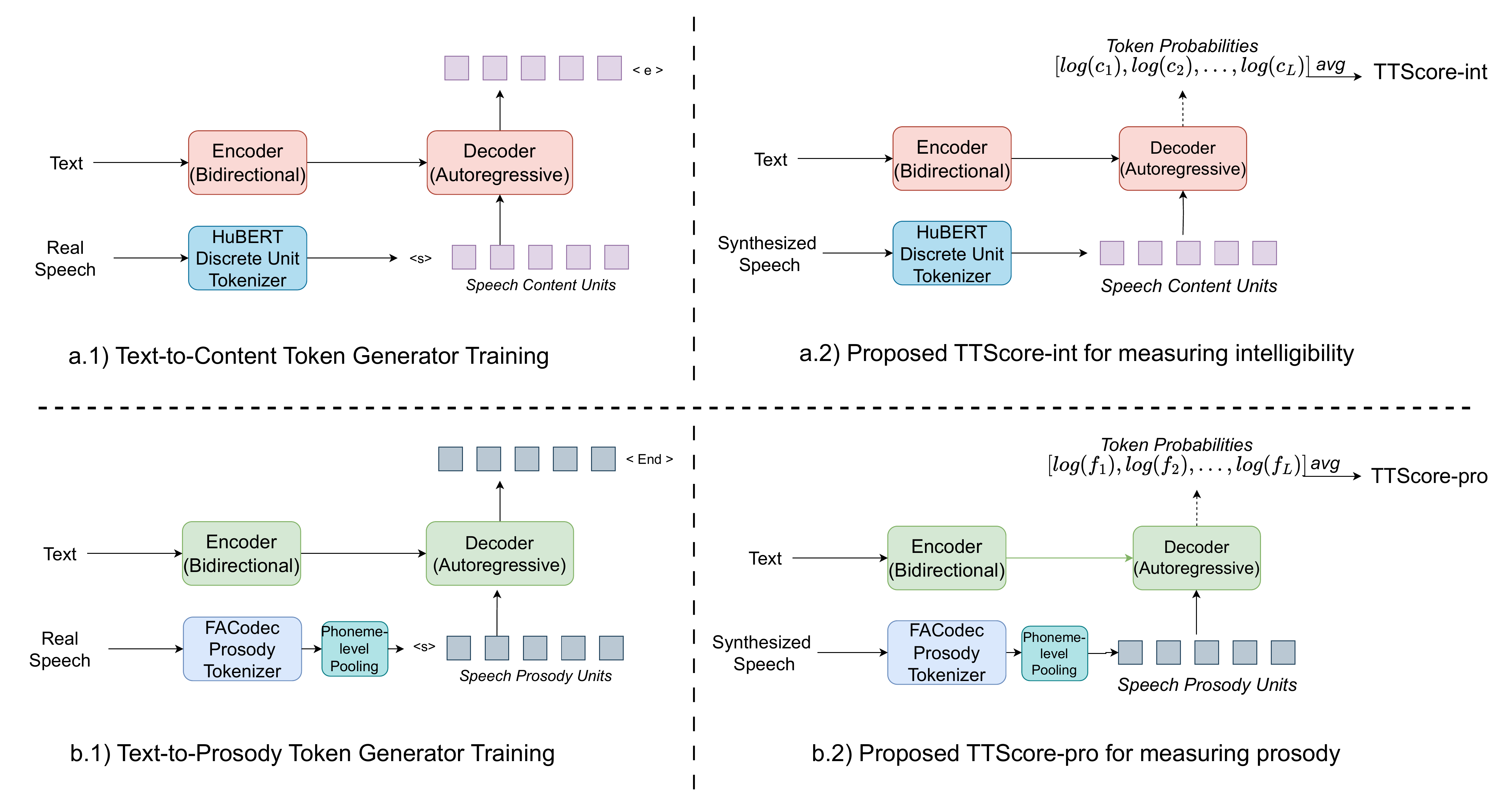}}
    \caption{a.1) Text-to-content token generator training a.2) TTScore-int as the conditional likelihood from content token generator for the given content token sequence from the synthesized speech and corresponding text b.1) Text-to-prosody token generator training with phoneme-level pooling b.2) TTScore-pro as the conditional likelihood of phoneme-level prosody tokens for a given text and synthesized speech}
    \vspace{-2mm}
\end{figure*}
\subsection{CHOICE OF SPEECH TOKENS}
\subsubsection{CONTENT SPEECH TOKENS}
\label{sec:content_okens}
Central to our method is the selection of discrete speech tokens that serve as linguistic content representations for measuring intelligibility. The most common approach is to obtain tokens via \textit{k}-means quantization of intermediate representations from HuBERT~\cite{Hsu2021HuBERTSS}. Prior studies have shown that the dominant information captured by these discrete SSL tokens depends on both the choice of the intermediate layer and the number of clusters (\textit{k}) used in the quantization process. Earlier HuBERT layers tend to encode acoustic and paralinguistic information, while later layers increasingly capture linguistic content~\cite{pasad2023comparative,lin2023utility}. In this work, we explore different HuBERT layers and cluster sizes to identify effective configurations. The resulting discrete SSL tokens are used as \textit{content speech tokens}, which primarily encode the linguistic content of a given utterance and form the basis for our intelligibility metric.  
\vspace{-3mm}
\subsubsection{PROSODIC SPEECH TOKENS}

\label{sec:prosody_tokens}
For prosodic speech tokens, we adopt discrete prosody representations from FACodec \cite{naturalspeech3}, which has demonstrated strong prosody modeling and controllability in speech synthesis experiments with NaturalSpeech~3~\cite{naturalspeech3}. FACodec prosody tokens primarily encode F0 information, one of the most crucial indicators of prosody, making them well-suited for prosody evaluation. FACodec provides a learned, discretized, and normalized representation of pitch-related prosody, which reduces sensitivity to speaker-dependent pitch ranges, absolute F0 values, and recording conditions. FACodec employs residual vector quantization (RVQ) to factorize and disentangle speech into discrete representations corresponding to different attributes: prosody, content, and acoustic detail. To ensure effective disentanglement, explicit training objectives are imposed on each token type in addition to the speech reconstruction objective. For prosody tokens, these objectives include normalized F0 contour prediction and adversarial phone classification.  

FACodec prosody tokens, denoted as $\mathbf{f} = [f_1, f_2, ..., f_T]$, are generated at the frame level, where $T$ is the number of frames. Predicting such fine-grained prosody at the frame level is a complex task. Therefore, in this work, we refactor frame-level prosody tokens to phoneme-level representations to simplify the prediction task and better model prosody token distributions. For phoneme-level refactoring, first, we extract frame-level continuous prosody representations, $\mathbf{r} = [r_1, r_2, ..., r_T]$, before applying RVQ. Next, we align these frame-level features with phonemes, $\mathbf{x} = [x_1, x_2, ..., x_L]$, using timestamps obtained from a forced aligner, where $L$ is the number of phonemes. For each phoneme, we pool the corresponding frame-level continuous features as 
$r_{\text{ph}} = [ \text{pool}(r_{x1[\text{begin}]},..,r_{x1[\text{end}]}), \text{pool}(r_{x2[\text{begin}]},..,r_{x2[\text{end}]}), ...]$.

After applying RVQ to the phoneme-level pooled representations $r_{\text{ph}}$, we obtain the phoneme-level prosody tokens $\mathbf{f_{ph}} = [f_1, f_2, ..., f_L]$. In addition to simplifying the prosody prediction task, this downsampling from frame-level to phoneme-level also improves computational efficiency by significantly reducing sequence length.

\vspace{-3mm}
\subsection{TTSCORE DEFINITION \& METHODOLOGY}
\subsubsection{TEXT-TO-SPEECH TOKEN GENERATOR}
\label{sec:generators}
To estimate the distribution of discrete speech tokens in a synthesized utterance given its textual content, we train a sequence-to-sequence text-to-token generator that autoregressively predicts discrete speech tokens while incorporating content information from the input text. Predicting specialized and distilled speech tokens is expected to be a simpler problem than speech generation itself, as the generator only needs to model limited aspects of speech in a heavily compressed discrete token space, rather than the full set of details required to reconstruct a speech waveform. This task can be even simpler for intelligibility, which can be evaluated using semantic speech tokens (e.g., HuBERT units) that already exhibit strong correspondence with input phoneme classes \cite{pasad2023comparative}. For this reason, text-to-specialized token generators have strong potential to effectively model token distributions due to a simpler objective, and they also tend to be more interpretable due to their focus on specific speech attributes making them suitable candidates as evaluators.

As generator, we adopt a transformer-based encoder--decoder architecture, specifically BART~\cite{lewis-etal-2020-bart}, which has been widely used for various sequence-to-sequence tasks. We train two text-to-token generators for TTScore-int and TTScore-pro. Both generators takes a phoneme sequence, $\mathbf{x} = [x_1,x_2, ..., x_L]$, derived from the input text using a grapheme-to-phoneme converter, as the input. The difference between the two generator is the output tokens that they generate making them specialized in modeling different aspects of synthesized speech. The generator for TTScore-int predicts the content token sequence $\mathbf{c} = [c_1,c_2,...,c_T]$, for the given utterance, obtained using a pre-trained SSL model and k-means clustering as described in Section~\ref{sec:content_okens}. The generator for TTScore-pro predicts phoneme-level prosody token sequence $\mathbf{f_{ph}}=[f_1,f_2,...,f_L]$ described in Section~\ref{sec:prosody_tokens}. $L$ and $T$ denote lengths of phoneme sequence and discrete speech token sequence, respectively. Both models are trained on a large-scale dataset of paired text and speech token sequences, enabling them to learn the distributions of aspect-specific tokens in natural speech.  

\vspace{-4mm}
\subsubsection{TTScore}
\label{sec:ttscore}
The overall framework for the text-to-token generators and likelihood scoring is illustrated in Figure 3.  
Using the seq2seq text-to-token generators described in Section~\ref{sec:generators}, with parameters $\theta_c$ for the content token generator and $\theta_f$ for the prosody token generator, we formulate the conditional probability of content tokens $\mathbf{c}$ and prosody tokens $\mathbf{f}_{\text{ph}}$ given the input phoneme sequence $\mathbf{x}$ as  
\setlength{\abovedisplayskip}{4pt}
\setlength{\belowdisplayskip}{4pt}
\begin{align}
\small
        p(\mathbf{c}|\theta_c,\mathbf{x}) = \prod p(c_i|\mathbf{c}_{<i},\mathbf{x}) \\
\small        p(\mathbf{f_{ph}}|\theta_f,\mathbf{x}) = \prod p(f_j|\mathbf{f}_{<j},\mathbf{x}) 
\end{align}

where $i\in\{1,2,...,T\}$ and $j\in\{1,2,...,L\}$.

This joint probability represents the expectation of the discrete speech token sequence that should occur for the corresponding content. For a given synthesized speech, if the content token sequence is likely for the given input text according to content token generator, we assume better intelligibility as the sequence contains the expected linguistic information. Similarly if the prosody token sequence is likely for the input text according to the prosody token generator, we assume a more appropriate pitch-related prosody for the given linguistic information. 

These probabilities can be refactored to a log-probability scores which constitues TTScore-int and TTScore-pro: 
\setlength{\abovedisplayskip}{3pt}
\setlength{\belowdisplayskip}{2pt}
\begin{align}
\small
    \text{TTScore-int} = P(\mathbf{c}|\mathbf{x},\theta_c) = \frac{1}{T} \sum_{i=1}^T \text{log}[ p(c_i|\mathbf{c}_{<i},\mathbf{x})] \\
\small    \text{TTScore-pro} = 
    P(\mathbf{f_{ph}}|\mathbf{x},\theta_f) = \frac{1}{L} \sum_{i=1}^L \text{log}[ p(f_i|\mathbf{f}_{<i},\mathbf{x})]
\end{align}

In order to measure intelligibility and prosody for a synthesized utterance, we first extract the corresponding discrete speech token sequences, content tokens $\mathbf{c} = [c_1, c_2, \ldots, c_T]$ and prosody tokens $\mathbf{f}_{\text{ph}} = [f_1, f_2, \ldots, f_L]$, using pre-trained speech tokenizers. The text-to-speech token generators, together with the conditional log-probability formulation, then evaluate these sequences by computing their probabilities as the average log-likelihood of each token, conditioned on the preceding tokens and the input phoneme sequence $\mathbf{x}$.

\section{EXPERIMENTS}
\subsection{DATASET}
For training both the text-to-content token generator and the text-to-prosody token generator, we use LibriSpeech-960~\cite{librispeech}, a large and diverse dataset containing 960 hours of speech from 2,311 speakers. This dataset is also used for training the HuBERT-base model~\cite{Hsu2021HuBERTSS}. To train the \textit{k}-means quantization models on top of the HuBERT features, we utilize the LibriSpeech-100 subset.  A reliable evaluation of both the proposed and baseline metrics requires a large number of synthesized samples from diverse systems, as well as extensive subjective ratings from human listeners. To meet these requirements, we evaluate our metric using the SOMOS~\cite{somos} and VoiceMOS~\cite{voicemos24,voicemos22} datasets, two widely used benchmarks that provide system diversity and large-scale MOS ratings.  SOMOS contains approximately 20,000 synthesized samples generated by 200 different neural network-based speech synthesis systems, each annotated with multiple MOS ratings of overall quality. For our experiments, we use the test-clean subset, consisting of about 3,000 utterances. VoiceMOS is the largest benchmark for automatic evaluation of speech generation. It includes samples from systems that participated in past Voice Conversion Challenges (VCC)~\cite{vcc}, Blizzard Challenges~\cite{black2005blizzard}, and methods from the ESPNet toolkit~\cite{watanabe2018espnet}. We use the combined training and test subsets of the VoiceMOS22 main challenge, totaling 6,000 utterances.

\subsection{MODEL ARCHITECTURE \& IMPLEMENTATION DETAILS}
\label{sec:experimental}
For both text-to-speech token generators, we use a smaller version of the BART architecture~\cite{lewis-etal-2020-bart}\footnote{\url{https://github.com/neulab/BARTScore}}. Our implementation consists of 6 transformer encoder layers and 6 transformer decoder layers. Each layer has dimension 512, and input/output embedding dimensions are 256. The model uses 8-head multi-head attention for both encoder and decoder. Training is performed with a dropout rate of 0.1, the AdamW optimizer, and a batch size of 8. The maximum sequence length is set to 1024 for efficiency.  

For speech content tokens, we use the pre-trained HuBERT-base model with 12 transformer encoder layers. We experiment with discrete tokens extracted from different HuBERT layers (3, 9, 12) and with different cluster sizes (\textit{k}=50, 500) for \textit{k}-means. For prosody tokens, we use the pre-trained FACodec tokenizer\footnote{\url{https://github.com/lifeiteng/naturalspeech3_facodec}}, which operates at a 12.5 ms frame rate with a codebook size of 1024. Phoneme boundaries are obtained using the Montreal Forced Aligner~\cite{mfa}\footnote{\url{https://github.com/MontrealCorpusTools/Montreal-Forced-Aligner}} and official transcripts. Phoneme-level prosody tokens are then computed as described in Section~\ref{sec:prosody_tokens}.

\subsection{EVALUATION SETUP FOR INTELLIGIBILITY}
We compare our proposed TTScore-int against the following baselines:

\textbf{WER and CER}: The standard intelligibility metrics. Synthesized speech is transcribed with a state-of-the-art ASR system (wav2vec 2.0\footnote{\url{https://huggingface.co/facebook/wav2vec2-large-960h-lv60-self}}), and WER/CER are computed against ground-truth transcripts.  

 \textbf{SpeechLMScore}~\cite{Maiti2023SpeechlmscoreES}: A likelihood-based method that models autoregressive generation of discrete speech tokens without text input\footnote{\url{https://github.com/soumimaiti/speechlmscore_tool}}.  
 
 \textbf{uLM:} For a fair comparison, we also implement a discrete speech token language model (uLM) under the same settings as our model. uLM is a decoder-only BART variant (Section~\ref{sec:experimental}) that autoregressively predicts tokens without text input.  


To assess how well the proposed metric reflects speech intelligibility, we measure correlations between the proposed metric and standard WER/CER for synthesized speech samples. Although WER and CER are imperfect, they are reliable and widely accepted objective metrics for intelligibility, we assume that a high correlation with them is a fundamental indicator of measuring objective intelligibility and a requirement for an objective intelligibility metric. We use Pearson linear correlation coefficient (LCC)\cite{pearson} to assess the correlations on both utterance level and system level where the correlation calculated between utterance scores and averaged sytem scores, respectively. Most importantly, we evaluate how well the proposed metric reflects the perceived quality of synthesized speech by measuring correlations with human MOS ratings. In this evaluation, we compare the proposed metric with WER and CER to investigate whether measuring intelligibility in the speech domain better reflects human perception of speech.  We report LCC and Spearman rank correlation coefficient (SRCC)\cite{Spearman2015ThePA} on both utterance level and system level.
\subsection{EVALUATION SETUP FOR PROSODY}

\subsubsection{REAL VS SYNTHESIZED SPEECH SCORES}
Firstly, we perform a score distribution analysis of proposed prosody metric, TTScore-pro, by comparing real and synthesized speech to validate the sanity of prosody evaluation. We assume that real speech exhibits more natural prosody compared to synthesized speech, which has limited prosody modeling capabilities. We hypothesize that a reliable prosody consistency metric should produce higher scores for real speech as a fundamental requirement. For real speech, we calculated scores for utterances from the LibriSpeech-dev~\cite{librispeech} and VCTK~\cite{VCTK}, which contains diverse content and speaking styles. For synthesized speech, we used scores from the VoiceMOS and SOMOS datasets, as both cover a wide range of synthesis methods.
\vspace{-5mm}
\subsubsection{ORIGINAL VS PERTURBED PITCH SCORES}
In order to see if the proposed score, TTScore-pro, is sensitive to prosody perturbations and accounts for prosody appropriateness, we perform pitch-controllable resynthesis experiments.  We performed resynthesis experiments on two different datasets, HifiTTS~\cite{hifitts} and librispeech-dev.  We utilize a SOTA speech analysis-synthesis method to resynthesize speech with the original pitch sequence of the utterance and a perturbed pitch contour.  We hypothesize that a reliable prosody metric should be sensitive and account for unnatural conditions of the resynthesized speech with a perturbed pitch. We utilize NANSY++~\cite{choi2023nansy} as the resynthesis method and apply the following pitch perturbations:

 \textbf{Original F0:} No perturbation is applied.
 
 \textbf{Inverse F0:} We invert f0 values around the mean f0 value to disrupt the f0 contour. This perturbation flips the contour around the mean and reverses the f0 increase/decrease events. 
 
 \textbf{Flipped F0:} We flip the f0 contour with respect to the time axis, resulting in a misaligned f0 contour with the content of the utterance.

A reliable prosody metric should yield lower scores for perturbed conditions than for original F0.  
\vspace{-2mm}
\subsubsection{CORRELATION WITH MOS AND TTSArena ELO}
Since prosody is a core component of perceived naturalness, we further evaluate TTScore-pro by measuring correlations with human MOS ratings from SOMOS~\cite{somos} and VoiceMOS~\cite{voicemos24}. Both LCC and SRCC are reported at utterance and system levels.  

We also compare TTScore-pro with conventional metrics: 
\begin{itemize}
    \item \textbf{F0-RMSE}, which measures root mean-square error between synthesized and reference F0 contours \cite{Clark1999ObjectiveMF}. 
    \item \textbf{F0-correlation}, which measures Pearson correlation between synthesized and reference contours \cite{Clark1999ObjectiveMF}.
\end{itemize}
 
Finally, we evaluate correlation with ELO scores from the TTS Arena platform~\cite{tts-arena}, which are derived from large-scale pairwise preference tests. We use 535 synthesized samples from 10 state-of-the-art TTS systems. A reliable prosody metric should align with these perceptual judgments.  

\begin{table*}[!th]
\centering
\caption{Correlations of objective metrics with WER and CER in SOMOS and VoiceMOS datasets. (with bootstrapped 95\% confidence intervals)}
\label{table:wer}
\scalebox{0.68}{
\begin{tabular}{c|ccc|cccc|cccc}
\hline
\multirow{3}{*}{Model} & \multicolumn{3}{c|}{\multirow{2}{*}{Discrete Speech token}} & \multicolumn{4}{c|}{SOMOS Dataset} & \multicolumn{4}{c}{VoiceMOS22 Dataset} \\
\cline{5-12}
& & & & \multicolumn{2}{c}{WER} & \multicolumn{2}{c}{CER} & \multicolumn{2}{c}{WER} & \multicolumn{2}{c}{CER} \\
\hline
& Model& Layer& k&  Utt-level  & Sys-level & Utt-level  & Sys-level & Utt-level  & Sys-level & Utt-level  & Sys-level \\
\hline
SpeechLMScore & HuBERT & L3 & 50 & 0.08 (0.05, 0.12) & 0.22 (0.09, 0.35) & 0.11 (0.07, 0.15) & 0.21 (0.07, 0.34) & 0.32 (0.30, 0.34) & 0.42 (0.28, 0.54) & 0.31 (0.29, 0.33) & 0.40 (0.26, 0.52) \\
\hline
\multirow{6}{*}{uLM (ours)} & \multirow{6}{*}{HuBERT} 
& L3 & 50 & 0.08 (0.05, 0.12) & 0.21 (0.07, 0.34) & 0.07 (0.03, 0.10) & 0.16 (0.02, 0.29) & 0.30 (0.28, 0.32) & 0.39 (0.27, 0.53) & 0.30 (0.28, 0.32) & 0.38 (0.25, 0.52) \\
& & L3 & 500 & 0.16 (0.13, 0.20) & 0.39 (0.28, 0.51) & 0.19 (0.16, 0.23) & 0.39 (0.28, 0.52) & 0.33 (0.31, 0.35) & 0.45 (0.33, 0.57) & 0.32 (0.30, 0.34) & 0.44 (0.32, 0.58) \\
& & L9 & 50 & 0.24 (0.21, 0.27) & 0.43 (0.32, 0.55) & 0.21 (0.18, 0.25) & 0.37 (0.26, 0.49) & 0.33 (0.31, 0.35) & 0.44 (0.32, 0.57) & 0.33 (0.31, 0.35) & 0.43 (0.31, 0.57) \\
& & L9 & 500 & 0.42 (0.39, 0.45) & 0.69 (0.63, 0.76) & 0.43 (0.40, 0.46) & 0.68 (0.59, 0.76) & 0.44 (0.42, 0.46) & 0.55 (0.45, 0.67) & 0.45 (0.43, 0.47) & 0.57 (0.48, 0.68) \\
& & L12 & 50 & 0.20 (0.17, 0.24) & 0.37 (0.26, 0.49) & 0.19 (0.15, 0.23) & 0.33 (0.21, 0.45) & 0.32 (0.30, 0.34) & 0.41 (0.29, 0.55) & 0.32 (0.30, 0.34) & 0.40 (0.28, 0.54) \\
& & L12 & 500 & 0.32 (0.29, 0.35) & 0.54 (0.45, 0.65) & 0.34 (0.31, 0.37) & 0.55 (0.46, 0.65) & 0.41 (0.39, 0.43) & 0.51 (0.40, 0.63) & 0.37 (0.35, 0.39) & 0.51 (0.40, 0.63) \\
\hline
\multirow{6}{*}{\textbf{TTScore-int(proposed)}} & \multirow{6}{*}{HuBERT} 
& L3 & 50 & 0.38 (0.35, 0.41) & 0.67 (0.50, 0.76) & 0.38 (0.35, 0.41) & 0.64 (0.57, 0.73) & 0.77 (0.76, 0.78) & 0.94 (0.92, 0.96) & 0.76 (0.75, 0.77) & 0.94 (0.92, 0.96) \\
& & L3 & 500 & 0.37 (0.34, 0.40) & 0.70 (0.64, 0.78) & 0.43 (0.40, 0.46) & 0.71 (0.65, 0.79) & 0.75 (0.74, 0.76) & 0.93 (0.91, 0.95) & 0.74 (0.73, 0.75) & 0.96 (0.94, 0.97) \\
& & L9 & 50 & 0.48 (0.45, 0.51) & 0.77 (0.72, 0.83) & 0.50 (0.47, 0.53) & 0.73 (0.67, 0.80) & 0.78 (0.77, 0.79) & 0.94 (0.92, 0.96) & 0.78 (0.77, 0.79) & 0.95 (0.94, 0.97) \\
& & L9 & 500 & \textbf{0.53 (0.51, 0.56)} & \textbf{0.78 (0.73, 0.84)} & \textbf{0.53 (0.51, 0.56)} & \textbf{0.77 (0.72, 0.83)} & 0.74 (0.73, 0.76) & 0.95 (0.94, 0.97) & 0.74 (0.73, 0.75) & 0.96 (0.95, 0.98) \\
& & L12 & 50 & 0.47 (0.44, 0.50) & 0.75 (0.70, 0.82) & 0.50 (0.47, 0.53) & 0.73 (0.67, 0.80) & 0.77 (0.76, 0.78) & 0.94 (0.92, 0.96) & 0.77 (0.76, 0.78) & 0.95 (0.94, 0.97) \\
& & L12 & 500 & 0.48 (0.46, 0.51) & 0.76 (0.71, 0.83) & 0.53 (0.50, 0.56) & 0.77 (0.72, 0.83) & \textbf{0.77 (0.76, 0.78)} & \textbf{0.96 (0.95, 0.98)} & \textbf{0.77 (0.76, 0.78)} & \textbf{0.97 (0.96, 0.98)} \\
\hline
\end{tabular}}
\vspace{-2mm}
\end{table*}

\vspace{-2mm}

\begin{table*}[!h]
     \centering
     
     \caption{Correlations of objective metrics with MOS in SOMOS and VoiceMOS datasets. (with bootstrapped 95\% confidence intervals)}

     \label{table:mos_int}
     \scalebox{0.66}{
     \setlength{\tabcolsep}{3pt}
     \begin{tabular}{ccccccc|cccc|cc}
        \hline
         \multirow{3}{*}{Metric }&\multicolumn{2}{c}{\multirow{3}{*}{Speech token} } & \multicolumn{4}{c|}{SOMOS } & \multicolumn{4}{c|}{VoiceMOS22 (all) } & \multicolumn{2}{c}{VoiceMOS22 (test) }\\
         \cline{4-13}
         & & & \multicolumn{2}{c}{Utt-level } & \multicolumn{2}{c|}{Sys-level } & \multicolumn{2}{c}{Utt-level } & \multicolumn{2}{c|}{Sys-level } & Utt-level & Sys-level \\
         & & & LCC & SRCC & LCC & SRCC & LCC & SRCC & LCC & SRCC & LCC & LCC\\
         \hline
         UTMOS \cite{saeki22c_interspeech} & - & - & 0.36 (0.30, 0.41) & 0.33 (0.27, 0.38) & 0.53 (0.42, 0.62) & 0.57 (0.46, 0.66) & - & - & - & - & \textbf{0.88 (0.87, 0.89)} & \textbf{0.94 (0.92, 0.96)} \\
         \hline
         \textit{WER} & \multicolumn{2}{c}{Word}& \textit{0.32 (0.29, 0.35)} & \textit{0.30 (0.27, 0.33)} & \textit{0.62 (0.54, 0.72)} & \textit{0.63 (0.55, 0.72)} & \textit{0.30 (0.28, 0.32)} & \textit{0.31 (0.29, 0.33)} & \textit{0.42 (0.30, 0.56)} & \textit{0.41 (0.27, 0.56)} & \textit{0.25 (0.19, 0.31)} & \textit{0.39 (0.27, 0.53)} \\ \textit{CER} & \multicolumn{2}{c}{Character} & \textit{0.34 (0.31, 0.37)} & \textit{0.33 (0.30, 0.36)} & \textit{0.60 (0.52, 0.70)} & \textit{0.68 (0.60, 0.76)} & \textit{0.28 (0.26, 0.30)} & \textit{0.31 (0.29, 0.33)} & \textit{0.40 (0.28, 0.54)} & \textit{0.48 (0.35, 0.62)} & \textit{0.23 (0.17, 0.29)} & \textit{0.38 (0.25, 0.52)} \\
         \hline
         \hline
         SpeechLMScore~\cite{Maiti2023SpeechlmscoreES} & {\makecell{HuBERT \\ (k=50)}}  & L3 & 0.04 (0.00, 0.08) & 0.04 (0.00, 0.08) & 0.12 (-0.02, 0.26) & 0.08 (-0.05, 0.22) & 0.48 (0.46, 0.50) & 0.50 (0.48, 0.52) & 0.56 (0.46, 0.68) & 0.65 (0.56, 0.76)  & 0.47 (0.42, 0.52) & 0.58 (0.48, 0.69)\\
         \hline
         uLM(ours) & {\makecell{HuBERT \\ (k=50)}}  & L3 & 0.06 (0.03, 0.09) & 0.06 (0.02, 0.08) & 0.17 (0.05, 0.30) & 0.14 (0.01, 0.28) & 0.46 (0.44, 0.48) & 0.46 (0.44, 0.48) & 0.53 (0.41, 0.63) & 0.63 (0.52, 0.73) & 0.45 (0.40, 0.50) & 0.55 (0.44, 0.65) \\

         \multirow{3}{*}{uLM (ours)} & \multirow{3}{*}{\makecell{HuBERT \\ (k=500)}} & L3 & 0.17 (0.13, 0.21) & 0.16 (0.12,0.20) & 0.37 (0.24, 0.48) & 0.36 (0.24, 0.47) & \textbf{0.57 (0.55, 0.59)} & \textbf{0.57 (0.55, 0.59)} & \textbf{0.69 (0.60, 0.76)} & \textbf{0.72 (0.62, 0.79)} & 0.59 (0.55, 0.63) & 0.66 (0.56, 0.74) \\
         & & L9 & 0.34 (0.31,0.37) & 0.33 (0.30, 0.36) & 0.62 (0.53, 0.70) & 0.62 (0.53, 0.70) & 0.47 (0.45, 0.49) & 0.48 (0.46, 0.50) & 0.61 (0.50, 0.70) & 0.63 (0.53, 0.72) & 0.47 (0.42, 0.52) & 0.56 (0.45, 0.66)\\
         & & L12 & 0.24 (0.21, 0.27) & 0.22 (0.18, 0.25) & 0.50 (0.39, 0.60) & 0.50 (0.40, 0.60) & 0.44 (0.42, 0.46) & 0.44 (0.42, 0.46) & 0.54 (0.42, 0.64) & 0.56 (0.45, 0.66) & 0.43 (0.38, 0.48) & 0.51 (0.38, 0.62) \\
        \hline
\textit{TTScore-int (proposed)} & {\makecell{HuBERT \\ (k=50)}}  & L3 & 0.37 (0.34, 0.40) & 0.35 (0.32, 0.38) & 0.64 (0.55, 0.72) & 0.60 (0.49, 0.70) & 0.42 (0.40, 0.44) & 0.47 (0.45, 0.49) & 0.48 (0.35, 0.59) & 0.56 (0.44, 0.69) & 0.40 (0.35, 0.45) & 0.48 (0.35, 0.59) \\

         \multirow{3}{*}{\textit{TTScore-int (proposed)}} & \multirow{3}{*}{\makecell{HuBERT \\ (k=500)}} & L3 & 0.40 (0.37, 0.43) & 0.38 (0.35, 0.41) & 0.69 (0.61, 0.76) & 0.68 (0.58, 0.74) & 0.48 (0.46, 0.50) & 0.52 (0.50, 0.54) & 0.54 (0.42, 0.64) & 0.64 (0.52, 0.74) & 0.47 (0.42, 0.52) & 0.54 (0.42, 0.64) \\
        & & \textbf{L9} & \textbf{0.43 (0.40, 0.46)} & \textbf{0.42 (0.39, 0.45)} & \textbf{0.72 (0.64, 0.78)} & \textbf{0.71 (0.63, 0.78)} & 0.42 (0.40, 0.44) & 0.46 (0.44, 0.48) & 0.50 (0.37, 0.61) & 0.58 (0.45, 0.69) & 0.41 (0.36, 0.46) & 0.51 (0.38, 0.62) \\
         & & L12 & 0.42 (0.39, 0.45) & 0.40 (0.37, 0.43) & 0.71 (0.64, 0.77) & 0.70 (0.62, 0.77) & 0.41 (0.39, 0.43) & 0.44 (0.42, 0.46) & 0.48 (0.35, 0.59) & 0.56 (0.44, 0.67) & 0.38 (0.32, 0.44) & 0.47 (0.34, 0.59) \\
        \hline

     \end{tabular}}
    \vspace{-2mm}
 \end{table*}

\section{RESULTS FOR INTELLIGIBILITY EVALUATION}

\subsection{CORRELATION WITH WER AND CER}
Table~\ref{table:wer} shows that the proposed conditional likelihood metric achieves very high correlations with WER and CER in both datasets, especially VoiceMOS, confirming its effectiveness for measuring intelligibility. Assessing speech tokens as a conditional generation task (TTScore-int) yields much stronger correlations than language modeling approaches (uLM, SpeechLMScore), since conditioning on text guides the predictor to capture more linguistic content more directly.

For HuBERT-based tokens, correlations with WER and CER increase with higher layers, consistent with the expectation that later layers contain more refined content information~\cite{Hsu2021HuBERTSS}. This layer effect is most visible in language modeling, where performance depends heavily on token representation. In contrast, TTScore-int benefits from textual conditioning, allowing it to capture semantic content reliably even with earlier-layer tokens, where para-linguistic information is also prominent.

The number of tokens has little effect on TTScore-int, though larger vocabularies generally correlate slightly better with WER and CER. In contrast, token language models are far more sensitive to vocabulary size, sometimes showing very low correlations. This again highlights the importance of text conditioning in producing a robust, semantically grounded intelligibility metric.


\subsection{CORRELATION WITH MOS}

Table~\ref{table:mos_int} reports correlations with MOS. WER and CER show moderate alignment with human ratings, confirming intelligibility’s importance for the perceived quality. TTScore-int achieves consistently stronger correlations, particularly in SOMOS, indicating that evaluating intelligibility in the speech domain with a fine-grained generative formulation better reflects human perception.

There are also differences between benchmarks. In VoiceMOS, where older systems suffer from issues such as robotic prosody and noise, factors beyond intelligibility may play a larger role, making general-purpose models like uLM slightly more competitive. In contrast, SOMOS contains more recent, higher-quality systems where intelligibility differences are more salient, and TTScore-int clearly outperforms baselines in both scenarios. Moreover, consistently high correlations in different extensive benchmarks indicate the generalization ability and robustness of the proposed metric. 

Additionally, we include UTMOS \cite{saeki22c_interspeech}, the top-performing system from the VoiceMOS 2022 Challenge, as a reference. Results are reported on the VoiceMOS test subset, as UTMOS is trained on the VoiceMOS 22 training data. As expected, UTMOS achieves high correlation with MOS on VoiceMOS, but its performance is notably worse on the SOMOS dataset. Although UTMOS is trained with MOS-labeled data and directly targets MOS prediction, it shows lower correlation with MOS than the proposed metric on SOMOS, highlighting the challenge of cross-dataset generalization for MOS predictors that rely on MOS-labeled data, which is often limited in scale. In contrast, the proposed aspect-specific evaluation is not limited by the need for MOS-labeled data and exhibits more stable behavior across datasets, likely due to its more focused and simpler objective.

\begin{figure*}[!t] 
  \centering
  \begin{minipage}[b]{0.32\linewidth}
  \scalebox{0.85}{
\includegraphics[width=\linewidth]{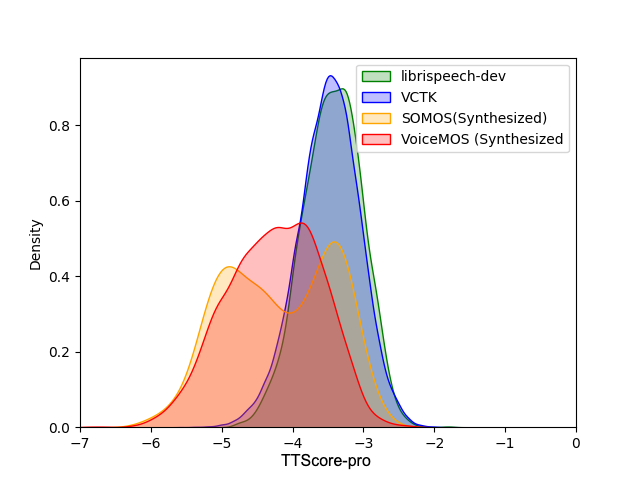}}
    \par(a) Score distributions for real vs synthesized speech
  \end{minipage}
  \begin{minipage}[b]{0.32\linewidth}
  \scalebox{0.85}{
    \includegraphics[width=\linewidth]{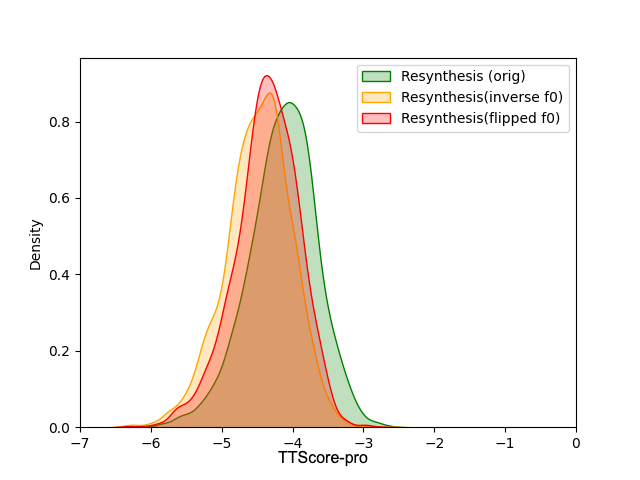}}
    \label{lab:tokenizers}
    \par(b) Score distributions with original vs perturbed pitch (HifiTTS)
  \end{minipage}
    \begin{minipage}[b]{0.32\linewidth}
    \scalebox{0.85}{
    \includegraphics[width=\linewidth]{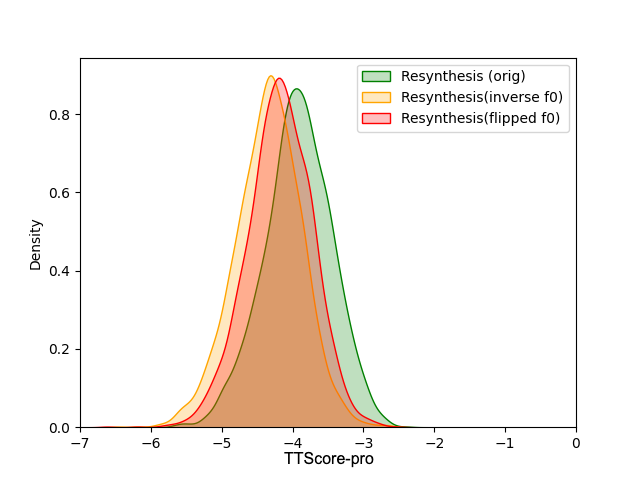}}
    \label{lab:tokenizers}
    \par(c) Score distributions with original vs perturbed pitch (librispeech-dev)
  \end{minipage}
  \caption{Score distribution analysis for the proposed prosody metric}

\end{figure*}

Overall, TTScore-int correlates more strongly with MOS than WER and CER across all conditions, demonstrating that text-conditioned, speech-domain evaluation captures both intelligibility and its contribution to overall quality more effectively than traditional text-domain metrics.

\subsection{DISCUSSION ON INTELLIGIBILITY \& QUALITY}

Our results demonstrate that intelligibility is closely correlated with MOS, particularly for more recent speech synthesis systems such as those in SOMOS. While intelligibility is a core component of speech perception and metrics like WER and CER are widely used to ensure it, their popularity is also due to the assumption that they partially reflect overall quality~\cite{alharthi2024evaluating}. These measures remain appealing because they provide a low-cost proxy for subjective evaluation, especially when large-scale human ratings are impractical.

There have also been recent efforts to predict MOS automatically~\cite{voicemos22,voicemos24}, but these approaches rely heavily on MOS-labeled data and often face issues of generalization and reliability~\cite{yamagishi_overview}.

Despite their widespread adoption, WER and CER have shortcomings as indicators of overall quality. A more accurate reflection of quality would provide deeper insight into synthesized speech and better guide system development. Our proposed method addresses these limitations, showing consistently stronger correlations with human ratings and offering a more informative alternative to conventional intelligibility metrics.






\section{RESULTS FOR PROSODY EVALUATION}

\subsection{REAL VS SYNTHESIZED SPEECH SCORES}

The score distributions of the proposed prosody metric for real and synthesized speech samples are shown in Figure 2(a). As observed in the figure, real speech samples with natural prosody achieve higher scores than synthesized speech samples with limited prosody modeling, supporting the validity and reliability of our metric. Furthermore, the similar scores across different datasets for both real and synthesized speech indicate the generalizability of our metric across different domains.
\begin{table}[!h]
\centering
\caption{Prosody correlations across benchmarks (with bootstrapped 95\% confidence intervals)}
\label{tab:prosody_all}
\setlength{\tabcolsep}{4.5pt}

\begin{minipage}{\linewidth}
\centering
\small\textit{(a) Correlation with MOS in SOMOS}
\vspace{2mm}
\scalebox{0.68}{
\begin{tabular}{c|cc|cc}
 & \multicolumn{2}{c|}{Utter.-level} & \multicolumn{2}{c}{System-level} \\
\hline
Metric & LCC & SRCC & LCC & SRCC \\
\hline
\textit{Log f0 RMSE} & \textit{-0.03 (-0.09, 0.03)} & \textit{-0.03 (-0.09, 0.03)} & \textit{-0.08 (-0.22, 0.06)} & \textit{-0.09 (-0.23, 0.05)} \\
\textit{f0-corr}     & \textit{ 0.02 (-0.04, 0.08)} & \textit{0.04 (-0.02, 0.10)} & \textit{ 0.07 (-0.07, 0.21)} & \textit{0.02 (-0.11, 0.17)} \\
\hline\hline
TTScore-pro (small)        & 0.05 (0.00, 0.11) & 0.04 (-0.02, 0.10) & 0.23 (0.10, 0.35) & 0.15 (0.03, 0.26) \\
\textbf{TTScore-pro (large)} & \textbf{0.06 (0.00, 0.12)} & \textbf{0.05 (-0.01, 0.11)} & \textbf{0.25 (0.11, 0.38)} & \textbf{0.18 (0.04, 0.32)} \\
\end{tabular}}
\end{minipage}

\vspace{3mm}
\begin{minipage}{\linewidth}
\centering
\small\textit{(b) Correlation with MOS in VoiceMOS}
\vspace{2mm}
\scalebox{0.75}{
\begin{tabular}{c|cc|cc}
 & \multicolumn{2}{c|}{Utter.-level} & \multicolumn{2}{c}{System-level} \\
\hline
Test Set & LCC & SRCC & LCC & SRCC \\
\hline
All Systems      & 0.33 (0.31, 0.35) & 0.33 (0.31, 0.36) & 0.45 (0.32, 0.57) & 0.46 (0.31, 0.60) \\
Top 25\% Systems & 0.20 (0.14, 0.26) & 0.21 (0.15, 0.27) & 0.25 (0.13, 0.34) & 0.32 (0.21, 0.41) \\
Top 12\% Systems & 0.19 (0.10, 0.28) & 0.19 (0.10, 0.28) & 0.30 (0.15, 0.44) & 0.24 (0.10, 0.39) \\
\end{tabular}}
\end{minipage}

\vspace{3mm}
\begin{minipage}{\linewidth}
\centering
\small\textit{(c) Correlation with TTSArena ELO}
\vspace{2mm}
\scalebox{0.92}{
\begin{tabular}{c|cc}
 & \multicolumn{2}{c}{System-level} \\
Metric & LCC & SRCC \\
\hline
\textit{Log f0 RMSE} & \textit{0.18* (-0.51, 0.73)} & \textit{0.32* (-0.40, 0.86)} \\
\textit{f0-corr}     & \textit{0.13 (-0.55, 0.70)}  & \textbf{\textit{0.32 (-0.36, 0.89)}} \\
\hline
\textbf{TTScore-pro (large)} & 0.25 (-0.45, 0.76) & 0.21 (-0.50, 0.73)\\
\end{tabular}}
\vspace{2mm}\\[-2pt]\footnotesize *Contradicting sign of correlation value.
\vspace*{-5mm}
\end{minipage}

\end{table}

\vspace{-2mm}
\subsection{ORIGINAL VS PERTURBED PITCH SCORES}
The score distributions for resynthesized speech with correct and perturbed prosody in Figure 2(b) and 2(c) show that resynthesized speech with original prosody has higher scores than resynthesized speech with perturbed pitch, consistently in both scenarios. Since pitch contours being disturbed or not is the only varying factor between the cases, the score distribution analysis proves that the proposed prosody metric accounts for pitch appropriateness, showing lower values for disrupted pitch. 


We note that, compared to real speech scores in Figure 2(a), resynthesized speech with original prosody tends to have lower values. This is expected, as resynthesis method has its own limitations that reduce naturalness relative to original recordings. Some overlap also exists between the distributions of original and perturbed prosody, which is anticipated since prosody is inherently context-dependent and variable across utterances. For this reason, analyzing the shift in general score distributions provides more meaningful insight than relying on absolute values. Although the degree of distributional shift is difficult to quantify, the key trend is clear: speech with original prosody consistently receives higher scores than speech with perturbed pitch. This confirms that our metric accounts for prosodic degradation.
\vspace{-2mm}
\subsection{CORRELATION WITH MOS AND TTSArena ELO}

 Since lower F0-RMSE indicates better pitch-related prosody, a negative correlation with MOS is expected, while positive correlations are expected otherwise. As both baselines require reference speech, results are reported only on SOMOS, while VoiceMOS lacks aligned references. As shown in Table~\ref{tab:prosody_all}(a), the baselines show weak or no correlation with MOS, confirming their limited reliability for reflecting perceived naturalness.

In contrast, TTScore-pro achieves consistently stronger correlations. While moderate in absolute terms, this is expected since naturalness depends on multiple factors beyond prosody. The correlations are lower in general compared to TTScore-int, which can be attributed to intelligibility's more obvious effect on overall perception and higher complexity of prosody modeling. Results from VoiceMOS (Table~\ref{tab:prosody_all}(b)) further confirm its robustness, likely due to the dataset’s greater diversity in synthesis methods and prosody modeling quality. Overall, TTScore-pro provides a more reliable and informative objective prosody metric than traditional measures.

Table~\ref{tab:prosody_all}(c) shows correlations with TTSArena ELO scores, where higher ELO indicates better perceived quality. While F0-RMSE should correlate negatively with ELO, it unexpectedly shows a positive trend, confirming its unreliability as a prosody metric. F0-corr performs slightly better but remains inconsistent across datasets. In contrast, TTScore-pro consistently aligns with human preferences in this additional benchmark, the results indicate that it tends to capture the appropriateness of pitch patterns and tend to reflect prosodic naturalness more reliably. We note that this benchmark has fewer number of synthesis systems. Overall results highlight TTScore-pro as a more robust and informative objective metric, compared to traditional pitch-based measures which fail to reflect overall perception.

\vspace{-4mm}
\section{LIMITATIONS}
While we examine a very important aspect of prosody, namely F0, prosody is a broad concept that also encompasses other speech attributes such as rhythm, energy, and phrasing. In this work, we focus on F0 due to its importance and because it is the most commonly measured prosodic feature in speech synthesis. \\
Moreover, since token distributions are learned from data, our method is data-driven and may be susceptible to underlying biases. In this work, we utilize a large-scale English dataset recorded mostly under clean and controlled conditions. We focus on this setting, as such conditions are most widely used and commonly focused on in speech synthesis research and benchmarking. However, our method may have limitations in scenarios where recording conditions are less controlled (e.g., spontaneous speech or in-the-wild settings) and does not currently support other languages. We also note that, since our method does not require labels beyond text transcriptions, which also can be obtained automatically, it is directly applicable to other scenarios where such data is available.
\section{CONCLUSION}
In this work, we introduced a targeted evaluation framework for synthesized speech that measures intelligibility and prosody as distinct aspects of speech quality. We proposed two conditional likelihood–based metrics: TTScore-int, which evaluates how well the linguistic content of speech is preserved, and TTScore-pro, which assesses the appropriateness of pitch, a crucial aspect of prosody, relative to the intended content. By leveraging specialized discrete speech tokens and text-conditioned sequence-to-sequence predictors, both metrics provide reference-free, speech-domain evaluations that overcome many of the shortcomings of conventional approaches such as WER, CER, and F0-based measures.

Comprehensive experiments across SOMOS, VoiceMOS, and TTSArena benchmarks demonstrated that TTScore-int and TTScore-pro not only correlate strongly with established baselines but also align more closely with human judgments of naturalness. These results confirm the value of targeted evaluation: aspect-specific metrics provide finer insights while remaining predictive of overall perceived quality. This work opens new opportunities for extending targeted evaluation beyond intelligibility and prosody. The proposed paradigm of conditional generation for evaluation is highly flexible for new formulations for different aspects of speech. 

\bibliographystyle{IEEEtran}
\bibliography{merged_clean}

\end{document}